\documentclass[preprint2]{aastex}

\usepackage{epsf}

\newcommand{\mTs}{\mbox{$<T_{\mathrm s}>$}}
\newcommand{\Ts}{\mbox{$T_{\mathrm s}$}}
\newcommand{\Tk}{\mbox{$T_{\mathrm k}$}}
\newcommand{\NHI}{\mbox{$N_\mathrm{HI}$}}
\newcommand{\HI}{\mbox{H\,{\sc i}}}
\newcommand{\kms}{\mbox{km s$^{-1}$}}
\newcommand{\cms}{\mbox{cm$^{-2}$}}
\newcommand{\mgii}{\mbox{Mg\,{\sc ii}}}

\shorttitle{Low Redshift 21cm Absorbers}
\shortauthors{Lane \& Briggs}

\begin{document}

\title {Two New Low Redshift 21cm Absorbers}

\author{W.M. Lane}

\affil{Naval Research Lab, Code 7213, 4555 Overlook Ave. SW,
Washington, DC, 20375}

\email{lane@rsd.nrl.navy.mil}

\and

\author{F.H. Briggs}
\affil{Kapteyn Astronomical Institute, Postbus 800, NL-9700 AV
  Groningen, The Netherlands.}
\email{fbriggs@astro.rug.nl}

\begin{abstract}  

As part of a larger program to identify low redshift radio analogues
of the damped Lyman$-\alpha$ (DLA) absorbers seen in the spectra of
high redshift quasars, Westerbork Synthesis Radio Telescope (WSRT)
observations have discovered two new HI 21cm absorption lines at
$z=0.394$ and $z=0.437$ in the spectra of the radio sources B~0248+430
and B~1243-072 respectively.  These sightlines and redshifts were
selected for study on the basis of the previously known low ionization
absorption lines of \mgii, and neither has been observed in the
Ly$\alpha$ line.  The 21cm line observations provide information on
column densities, temperatures and kinematics of the thickest cold
neutral clouds in the absorbers.

\end{abstract}


\section{Introduction}

The identification of redshifted QSO absorption line systems with
\NHI$\geq 2 \times 10^{20}$ \cms, typically referred to as damped
Ly$\alpha$ (DLA) systems, is important because the DLA absorbers are
thought to be the progenitors of present day ordinary galaxies.  The
nature of these high neutral \HI\ column density systems is under
debate.  At high redshift, study of the absorbers is made difficult
by the dimness of the protogalactic systems, and much of the work is
theoretical (eg. Prochaska \& Wolfe 1998; Haehnelt, Steinmetz, \&
Rauch 1998).  At low redshift there are fewer absorbers known, and
recent observations find a variety of host galaxy types, including
amorphous low surface brightness galaxies as well as spirals and
compact objects, (eg. Steidel et al.  1995; Le Brun et al. 1997;
Bowen, Tripp \& Jenkins 2001; Turnshek et al. 2001).  Thus there is a
need to identify more local, low and moderate redshift examples for
detailed study.

One way to identify these systems is to select low-redshift absorbers
by the presence of another, more common, low-ionization absorption
line such as the \mgii\ $\lambda\lambda 2796,2803$ \AA\ doublet.  The
metal-line systems can then be observed in either the UV regime for
DLA absorption against QSO's with sufficient UV flux (eg. Rao \&
Turnshek 2000), and/or in the radio for \HI\ 21cm absorption if the
QSO is also a radio source (Lane et al. 2001, in prep).  Unlike the
saturated DLA absorption feature, the 21cm absorption profile can
provide information about the velocity and temperature of the
absorbing clouds (eg. Lane, Briggs, \& Smette 2000; Briggs, de Bruyn,
\& Vermeulen 2001).  The radio-emitting region in a QSO is extended,
so \HI 21cm observations at the appropriate angular resolution can
also provide information on the spatial distribution of the absorbing
gas by probing several sightlines to the background source.  If the
information on velocity structure and spatial distribution of the gas
is combined, it is possible to construct a reasonable model of the
intervening galaxy responsible for the absorption (eg. Briggs et
al. 2001).

The opacity of the 21cm line is inversely sensitive to the excitation
or spin temperature, \Ts, of the gas; lower \Ts\ gas will create a
more opaque absorption profile.  Under usual conditions in the
multi-temperature-phase interstellar medium (ISM) of the Milky Way,
\Ts\ is coupled to the kinetic temperature, \Tk, of cold-phase gas
\citep{dl90}, although the two temperatures begin to diverge in the
case of warm-phase ISM gas \citep{liszt01}.  As a result, the 21cm
line profile can be used to estimate the kinetic temperature of the
cold neutral absorbing gas to which it is most sensitive.  This can be
done either by comparison to the temperature-insensitive DLA
absorption line in the same system, or by using the width of each
individual component to set an upper limit to the kinetic temperature
of the gas (Lane et al. 2000; Kanekar, Ghosh, \& Chengalur 2001).

Currently, there are roughly 25 redshifted 21cm absorbers known.  Of
these roughly one-third have absorption redshift $z_{abs} \approx
z_{em}$, the QSO redshift, indicating that the absorption occurs in
the QSO host galaxy.  The rest are systems which intervene along the
sightline to a QSO.  In this paper we present high velocity resolution
radio spectra for two new intervening low-redshift \HI\ 21cm absorbers
identified in a recent survey (Lane 2000). Neither object has been
observed for Ly$\alpha$ absorption, but we infer that both have column
densities \NHI$\geq 2 \times 10^{20}$ \cms.

\section{Deriving the Column Density}

The neutral column density contained in a 21cm absorption line is
computed by taking the integral over the absorption profile:

\begin{equation}
\NHI\ = 1.8\times10^{18} ~\Ts\ \int \tau(v)~dv ~~\cms 
\label{eq:NHI}
\end{equation}
where \Ts\ is the spin temperature of the absorbing gas,
assumed to be constant for each cloud, and $v$ is the velocity in
\kms.  $\tau(v)$ is the optical depth of the line at velocity $v$, and
includes a factor $f$ to account for the fraction of the continuum
source covered by the absorber.

Thermal broadening of an absorption feature provides an independent
constraint on the kinetic temperature, $\Tk \approx \Ts$, of
the gas:

\begin{equation}
 T_{\rm k} \leq  21.855 \times \Delta v^2  ~~{\rm K}
\label{eq:Tks}
\end{equation}

where $\Delta v$ is the full-width half max (FWHM) velocity for atomic
Hydrogen measured in \kms.  Unfortunately, bulk kinematical motions
and turbulence in the absorbing \HI\ gas will also broaden the
absorption line, so this is not always a stringent constraint on the
gas temperature.

\section{The Data}

The 21cm spectra were obtained on 1999 September 19 (B 0248+430) and
2000 September 6 (B 1243-072) using the Westerbork Synthesis Radio
Telescope (WSRT)\footnote{The Westerbork Synthesis Radio Telescope is
operated by the Netherlands Foundation for Research in Astronomy
(NFRA/ASTRON) with support from the Netherlands Foundation for
Scientific Research(NWO).} with UHF-high receivers and the DZB
correlator.  Both observations were made as followups to previous and
more tentative detections from an extensive 21cm survey of \mgii\
absorbers by Lane (2000).

For B~0248+430, the total integration time was just under 4 hours.  A
bandwidth of 0.625 MHz centered at 1018.94 MHz is divided into 256
channels to provide a channel width of 2.44 kHz.  No on-line smoothing
was applied, giving a velocity resolution of 0.86 \kms.  Observations
of 3C 48 were used to calibrate the flux density scale and the
passband.  For B~1243-072, the total integration time was roughly 9.5
hours.  A bandwidth of 1.25 MHz and 128 channels centered at 988.6 MHz
give a velocity resolution of 3.6 \kms.  Observations of 3C 147 were
used to calibrate the flux and bandpass.

Using standard routines in AIPS, both data sets were self-calibrated
to line-free continuum maps.  After removing the continuum emission,
map cubes were made and spectra extracted at the positions of the
QSOs.  The final spectra, offset to indicate the measured continuum
flux, are shown in Figures ~\ref{fig:0248dzb}\ and ~\ref{fig:1243dzb}.

\section{B~0248+430, $z_{abs} = 0.3941$}

\begin{figure}[t!]
\centering
\epsfxsize=1.0\columnwidth
\epsfbox{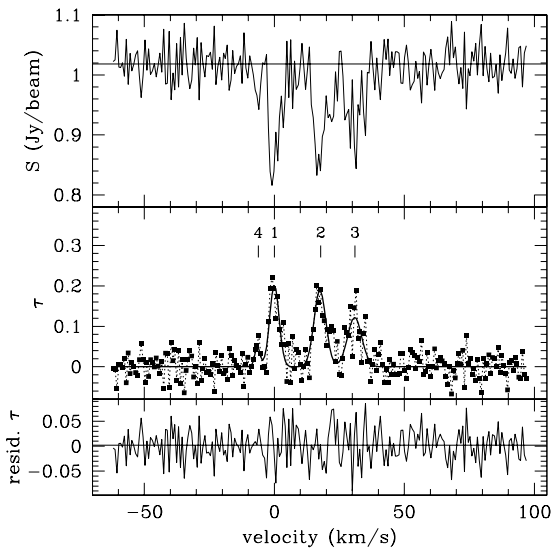}
\caption{\small WSRT/DZB high resolution spectrum of the $z = 0.3941$ \HI\
21cm absorber toward B~0248+430 at a channel spacing of 2.44 kHz, or a
velocity resolution of 0.81 \kms.  $v=0$ is placed at the metal-line
redshift of $z=0.394$.  The top panel shows the original spectrum.
The middle panel shows the optical depth of the lines, with the data
points connected by a light dotted line. The heavy solid line shows
the fit, and the centroids of the individual components listed in
Table 1 are marked.  The bottom panel shows residuals after the fit
has been subtracted. \label{fig:0248dzb}}
\end{figure}

\begin{table*}
\begin{center}
\small
\caption[Fit to the B 0248+430 21cm Absorption]{The Four
Component Fit to the B 0248+430 21cm line Profile}
\begin{tabular}[c]{*{6}{l}}
\tableline \tableline
Component & $\Delta$ V$_{\mathrm offset}$ & $\tau_{21}$ & FWHM
$\Delta v$ & $\int \tau(v)~dv$ & \Tk \\
~~ & (\kms) & ~~ & (\kms) & (\kms) & (K) \\ 
\tableline 
1 & 0 &  $0.20 \pm 0.03$ &  $4.3 \pm 0.4$ &  $0.92 \pm 0.16$ & $405
\pm 75$ \\
2 & $~17.8 \pm 0.6$ & $0.16 \pm 0.03$ &  $6.2 \pm 0.4$ & $1.05 \pm
0.21$ & $840 \pm 110$ \\
3 & $~31.0 \pm 0.6$ & $0.12 \pm 0.03$ & $6.9 \pm 0.4$ &  $0.88 \pm
0.23$ & $1040 \pm 120$ \\
4 &  $-6.2 \pm 0.6$ & $0.07 \pm 0.03$ &  $1.9 \pm 0.4$ &  $0.14 \pm
0.07$ & $79 \pm 33$ \\
\tableline
\end{tabular}
\end{center}
\label{tab:0248fit}
\end{table*}

B~0248+430 is a core dominated quasar at an emission redshift $z_{em}
= 1.31$.  The metal line system at $z_{abs} = 0.394$ was originally
identified in a spectrum taken to study absorption associated with a
pair of merging galaxies that lie close to the QSO sightline at $z =
0.05$ \citep{womble90}.  The initial detection of \mgii\ and Mg\,{\sc
i}\ at $z_{abs} = 0.394$ was confirmed, and Fe\,{\sc ii}\ and Ca\,{\sc
ii} absorption lines at the same redshift were reported by Sargent \&
Steidel (1990).  At present, there is no reported detection (or
significant non-detection) of the DLA line for this system.  For this
reason there is no temperature independent measure of the total
neutral column density in this system.

The \HI\ 21cm absorption feature is composed of a complex of lines,
roughly clustered in three groups, which combined cover $\sim 40$
\kms.  The $3\sigma$ noise in the spectrum is equivalent to an optical
depth, $\tau_{3\sigma} = 0.09$.  The best simultaneous 4-Gaussian
fit to the data is shown in Figure ~\ref{fig:0248dzb}, and the fitted
line parameters are listed in Table ~\ref{tab:0248fit}.  We have
placed the system velocity, $\Delta V = 0$, at a frequency of 1018.94
MHz, corresponding to both the center of the deepest absorption
component and the metal-line redshift of $z_{abs} = 0.394$.
Conservative errors for $\tau$ are estimated from the RMS of the
residuals.  The errors on $\Delta v$ are taken to be one-half of a
resolution element, and we assume that the gas covers the entire QSO
($f = 1$).

The 21cm line integral, $\int \tau(v)~dv = 2.99 \pm 0.36$ \kms, leads
to an estimated column density of \NHI$ = 5.4 \pm 0.6 \times 10^{18}
\times$ \mTs\ atoms \cms\ K$^{-1}$.  We use the notation \mTs\ to
indicate a column-density-weighted harmonic mean spin temperature for
all of the absorbing gas along the sightline.  Even if the spin
temperature of each component is as low as $\Ts\ \sim 100$ K, a
``typical'' value for cold-phase gas (Dickey \& Lockman, 1990), this
system would have a total $\NHI\ \approx 5 \times 10^{20}$ \cms.
Carilli et al. (1996) find $\mTs \approx 1000$ K to be more typical in
redshifted DLA/21cm absorbers, suggesting the true column density is
even higher.

The fitted absorption component widths are narrow enough to use Eq.
~\ref{eq:Tks} to calculate reasonable upper limits to \Tk\ for each.
These are listed in column 6 of Table ~\ref{tab:0248fit}.  Setting
$\Ts \leq \Tk$, we sum the derived column density for each of the 4
components and find a total column density $\NHI \leq 3.9 \pm 0.6
\times 10^{21}$ \cms\ is contained in the detected profile.  This
estimate is an upper limit because the lines may be broadened by bulk
kinematical motions as well as thermal motions.  On the other hand, we
could be excluding a significant column density of warm-phase \HI\
which would create a broad shallow absorption component not detected
at the sensitivity levels of this spectrum (Lane et al. 2000).

Optical and/or infrared followup of this absorber is made difficult by
the $z=0.05$ interacting galaxy pair, which includes a long tail that
crosses the QSO sightline.  Although many images of this field have
been published (eg. Borgeest et al. 1991, Kollatschny et al. 1991),
few candidate absorbers are known.  There is a small object roughly
1'' south of the QSO in an HST PC image (Maoz et al. 1992), and a
galaxy superimposed on the foreground tidal tail (Sargent \& Steidel
1990).  The closest two objects which are easily distinguished from
the tidal tail are a $z=0.240$ galaxy to the Northeast and a compact,
probably stellar object to the South (Womble et al. 1990).  It seems
unlikely that further ground-based imaging and spectroscopy will
identify the $z=0.394$ absorbing galaxy; space-based observations will
be necessary to identify candidate hosts against the extended low
surface brightness tidal tail.

At radio wavelengths, the background quasar can be resolved into a
double-lobed source with a separation along the northwest/southeast axis
of $\sim$12 mas at a frequency of 2 GHz (Fey \& Charlot 2000).
Mapping of the complex absorption against the two components will
provide information about the gas structure on a roughly 60 pc scale.

\section{B~1243$-$072, $z_{abs} = 0.4367$}

\begin{figure}[t!]
\centering
\epsfxsize=1.0\columnwidth
\epsfbox{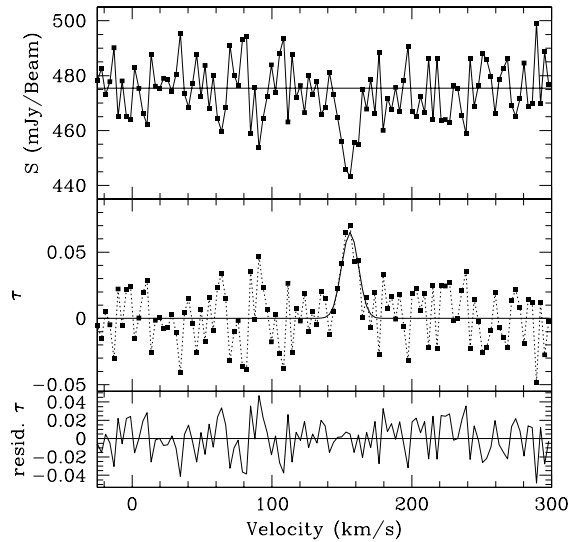}
\caption{\small WSRT/DZB spectrum of the $z = 0.4367$ \HI\ 21cm absorber
toward B~1243-072 at a channel spacing of 9.8 kHz, or a velocity
resolution of 3.6 \kms.  The velocity scale has been chosen such that
$v = 0$ corresponds to $z = 0.436$, the metal-line redshift.  The top
panel shows the original spectrum.  The middle panel shows the optical
depth of the line, with the data points connected by a light dotted
line, and a heavy solid line indicating the Gaussian curve we have fit
to the data.  The bottom panel shows residuals after the fit has been
subtracted. \label{fig:1243dzb}}
\end{figure}

B~1243-072 is a core-dominated quasar with an emission redshift
$z_{em} = 1.29$.  At 20cm it may have weak radio lobes extending
$\sim$ 5'' to the east and west (Gower \& Hutchings 1984; cf. Browne
\& Perley 1986); it is certainly unresolved by the $\sim$23''
resolution of the WSRT at this frequency.  The metal-line system at $z
= 0.436$ was originally identified by Wright et al. (1979), and
contains the \mgii\ $\lambda\lambda 2796, 2803$ doublet, and several
Fe\,{\sc ii}\ absorption features.  No measured equivalent widths are
reported for this early low-resolution spectrum; in fact the \mgii\
doublet is reported as a single line.  No subsequent spectra for this
source have been published.

The detected \HI\ 21cm absorption is shown in Figure
\ref{fig:1243dzb}.  The $3\sigma$ feature lies at a heliocentric
frequency of 988.60 MHz, corresponding to a redshift $z_{abs} =
0.4367$.  This is offset by some 160 \kms\ from the reported
metal-line redshift, probably due to the low precision of the optical
measurement.  The 21cm line is several channels wide and increases in
significance to $5\sigma$ after the spectrum is Hanning smoothed to 7
\kms\ velocity resolution.  A weak absorption feature has been present
at the same heliocentric frequency in two other, lower-quality, WSRT
spectra of this source (Lane 2000).

The absorption feature can be modelled as a single Gaussian component
with FWHM velocity $\Delta v = 14.1 \pm 1.8$ \kms\ and optical depth
$\tau = 0.065 \pm 0.020$, assuming a covering factor $f = 1$.  Errors
for $\tau$ are estimated from the RMS of the residuals, and the errors
on $\Delta v$ are taken to be one-half of a resolution element.  Using
these parameters we find a total \HI\ column density \NHI$ = 1.7 \pm
0.6 \times 10^{18} \times$ \mTs\ \cms.  We expect \mTs, the harmonic
mean spin temperature of all the gas in the absorption profile, to be
somewhere between the single cloud value $\Ts = 100$ K, and $\mTs =
1000$ K (eg. Carilli et al. 1996).  We conclude that this system
has a column density consistent with that of at least a weak DLA
absorber.

This system is ideal for detailed optical and radio follow-up study.
The quasar is only moderately bright (V = 18), making it relatively
easy to study sources quite close to the QSO sightline.  Currently,
the only published image of this field comes from the Digitized Sky
Survey (DSS)\footnote{The Digitized Sky Surveys were produced at the
Space Telescope Science Institute under U.S. Government grant NAG
W-2166.}, and the surrounding field appears relatively uncrowded.
There are only three or four very faint objects and one small bright
object within a projected radius of 25'' from the QSO sightline.  In
addition there is a large foreground disk galaxy at a projected
separation of about 1'.

Observations at a frequency of 2 GHz show a 20 mas extension to the
west of the radio quasar, making it a suitable candidate for very long
baseline interferometry (VLBI) mapping of the 21cm absorption (Fey \&
Charlot 2000).  Moderate baseline mapping might also be able to trace
absorption against the weak 5'' scale radio lobes (Gower \& Hutchings
1984).

\section{Conclusions}

We present two new \HI\ 21cm absorbers.  These add to a small but
growing group of high neutral column density systems at low and
moderate redshifts.  Both of these systems are suitable for further
detailed study which can contribute to our understanding of the types
of galaxies responsible for 21cm/DLA absorption and the evolution of
\HI-rich galaxies.

\acknowledgements

W. Lane is a National Research Council Postdoctoral Fellow.  Basic
research in astronomy at the Naval Research Laboratory is funded by
the Office of Naval Research.


\begin{thebibliography}{}

\bibitem[Borgeest et~al.(1991)]{b91}
Borgeest, U., Schramm, K.-J., Dietrich, M., Kollatschny, W., \& Hopp,
U. 1991, \aap, 243, 93

\bibitem[Bowen et~al.(2001)]{btj01}
Bowen, D. V., Tripp, T. M., \& Jenkins, E. B. 2001, \aj, 121, 1456

\bibitem[Briggs et~al.(2001)]{bdv01}
Briggs, F. H., de Bruyn, A. G., \& Vermeulen, R. C. 2001, \aap, 373, 113

\bibitem[Browne \& Perley(1986)]{bp86}
Browne, I.~W.~A. \& Perley, R.~A. 1986, \mnras, 222, 149

\bibitem[Carilli et~al.(1996)]{carilli96}
Carilli, C.~L., Lane, W., de Bruyn, A.~G., Braun, R., \& Miley,
  G.~K. 1996,  \aj, 111, 1830

\bibitem[Dickey \& Lockman(1990)]{dl90}
Dickey, J.~M. \& Lockman, F.~J. 1990,  \araa, 28, 215

\bibitem[Fey \& Charlot(2000)]{fc00}
Fey, A. L. \& Charlot, P. 2000, \apjs, 128, 17 

\bibitem[Gower \& Hutchings(1984)]{gh84}
Gower, A.~C. \& Hutchings, J. ~B. 1984, \aj, 89, 1658

\bibitem[Haehnelt et~al.(1998)]{hsr98}
Haehnelt, M. G., Steinmetz, M., \& Rauch M. 1998, \apj, 495, 647

\bibitem[Kollatschny et~al.(1991)]{k91}
Kollatschny, W., Dietrich, M., Borgeest, U., \& Schramm, K.-J. 1991,
\aap, 249, 57

\bibitem[Kanekar et~al.(2001)]{kgc01}
Kanekar, N. and Ghosh, T. \& Chengalur, J. N. 2001, \aap, 373, 394

\bibitem[Lane (2000)]{thesis}
Lane, W.M., 2000, Ph.D thesis, Univ. of Groningen

\bibitem[Lane et~al.(2000)]{lane00}
Lane, W., Briggs, F., \& Smette, A. 2000, \apj, 532, 146

\bibitem[LeBrun et~al.(1997)]{lebrun97}
Le Brun, V., Bergeron, J., Boisse, P., \& Deharveng, J. M. 1997, \aap, 321, 733

\bibitem[Liszt(2001)]{liszt01}
Liszt, H.  2001, \aap, accepted, astroph/0103246

\bibitem[Maoz et~al.(1992)]{m92}
Maoz, D., Bahcall, J. N., Doxsey, R., Schneider, D. P., Bahcall,
N. A., Lahav, O. \& Yanny, B. 1992, \apj, 394, 51

\bibitem[Prochaska \& Wolfe(1998)]{pw98}
Prochaska, J.X., \& Wolfe, A.M. 1998, \apj, 507, 113

\bibitem[Rao \& Turnshek(2000)]{rt00}
Rao, S.~M. \& Turnshek, D.~A. 2000,  \apjs, 130, 1

\bibitem[Sargent \& Steidel(1990)]{ss90}
Sargent, W. L.~W. \& Steidel, C.~C. 1990,  \apj, 359, L37

\bibitem[Steidel et al.(1995)]{s95}
Steidel, C.C., Bowen, D.V., Blades, J.C., \& Dickenson, M. 1995,
\apj, 440, L45

\bibitem[Turnshek et al.(2001)]{t01}
Turnshek, D.A., Rao, S., Nestor, D., Lane, W.M., Monier, E., Bergeron,
J., \& Smette, A. 2001, \apj, 553, 288

\bibitem[Womble et~al.(1990)]{womble90}
Womble, D.~S., Junkkarinen, V.~T., Cohen, R.~D., \& Burbidge, E.~M.
  1990,  \aj, 100, 1785

\bibitem[Wright et~al.(1979)]{wright97}
Wright, A.~E., Peterson, B.~A., Jauncey, D.~L., \& Condon, J.~J. 1979, \apj, 229, 73 

\end{thebibliography}
\end{document}